\documentclass[letterpaper,english]{article}
\usepackage[T1]{fontenc}
\usepackage[latin9]{inputenc}
\usepackage{color}
\usepackage{babel}
\usepackage{amsmath}
\usepackage{amssymb}
\usepackage{graphicx}
\usepackage[bookmarks=false,
 breaklinks=false,pdfborder={0 0 1},backref=false,colorlinks=true]
 {hyperref}
\hypersetup{
 allcolors=blue}

\makeatletter

\pdfpageheight\paperheight
\pdfpagewidth\paperwidth

\providecommand{\tabularnewline}{\\}

\usepackage{babel}

\usepackage{arxiv}
\usepackage{booktabs}
\usepackage{xcolor}
\newcommand{\D}{\mathrm{\,d}}

\author{
\textbf{John C. Boik}\textsuperscript{1},
\textbf{Kobus Esterhuysen}\textsuperscript{1,6},
\textbf{Jacqueline B. Hynes}\textsuperscript{1,2},
\textbf{Axel Constant}\textsuperscript{3},\\
\textbf{Ines Hipolito}\textsuperscript{1,5},
\textbf{Mahault Albarracin}\textsuperscript{1,4},
\textbf{Alex B. Kiefer}\textsuperscript{1},
\textbf{Karl Friston}\textsuperscript{1,7}
}

\date{
\textsuperscript{1}VERSES, Los Angeles, CA, USA\\
\textsuperscript{2}Spatial Web Foundation, Los Angeles, CA, USA\\
\textsuperscript{3}Dept. of Engineering and Informatics, Univ. of Sussex, Brighton, UK\\
\textsuperscript{4}Institut Sante et societe, UQAM, Montreal, CAN\\
\textsuperscript{5}Macquarie University, Sydney, AUS\\
\textsuperscript{6}LearnableLoopAI.com, Custer, WA, USA\\
\textsuperscript{7}Queen Square Institute of Neurology, UCL, London, UK
}

\makeatother

\begin{document}
\title{EcoNet: Multiagent Planning and Control of Household Energy Resources
Using Active Inference}
\maketitle
\begin{abstract}
Advances in automated systems afford new opportunities for intelligent
management of energy at household, local area, and utility scales.
Home Energy Management Systems (HEMS) can play a role by optimizing
the schedule and use of household energy devices and resources. One
challenge is that the goals of a household can be complex and conflicting.
For example, a household might wish to reduce energy costs and grid-associated
greenhouse gas emissions, yet keep room temperatures comfortable.
Another challenge is that an intelligent HEMS agent must make decisions
under uncertainty. An agent must plan actions into the future, but
weather and solar generation forecasts, for example, provide inherently
uncertain estimates of future conditions. This paper introduces EcoNet,
a Bayesian approach to household and neighborhood energy management
that is based on active inference. The aim is to improve energy management
and coordination, while accommodating uncertainties and taking into
account potentially conditional and conflicting goals and preferences.
Simulation results are presented and discussed.
\end{abstract}
\keywords{Active Inference\and Home Energy Management Systems\and Distributed Energy Resources\and Multi-Agent Systems\and Bayesian Inference\and Smart Grid}

\section{Introduction}

At any given moment, households and the utilities that supply them
with electricity face numerous choices regarding use and management
of their energy-related devices. Behind-the-meter devices can include
household heating, ventilation, and air conditioning (HVAC) systems
and distributed energy resources (DERs) such as solar panels and solar
and electric vehicle (EV) batteries. Utilities manage a wide assortment
of devices that also can include solar and battery systems. For example,
a utility might employ batteries to protect low-voltage transformers
\cite{RAOUFMOHAMED2021102975}.

The preferences for---and goals of---energy management can be conditional
and can conflict. A household might want to minimize energy costs,
yet wish to keep air conditioning systems on high during hot summer
afternoons, even during periods of peak energy costs, when time-of-use
(ToU) rates are high. A household might want to keep a room pleasantly
warm in winter, but only if the room is expected to be occupied, and
only if ToU rates are low. A utility might want to encourage home
charging of electric vehicle (EV) batteries, but to discourage large
numbers of households from doing so at the same moment. As a last
example, a household might want to charge a solar battery while the
sun is shining or ToU rates are low, but discharge when rates are
high. To ensure the battery has adequate charge to serve household
needs during high ToU periods, the household might need to track weather
forecasts and plan ahead, perhaps many hours in advance.

Intelligent energy management that adapts to custom goals and preferences
is important for comfort, safety, cost, and environmental impact,
including impacts related to energy-associated greenhouse gas emissions.
DER penetration is increasing \cite{osti_1784528}, in part due to
climate change concerns, and as it does there is growing need to coordinate
energy use and generation, not just within a household but also within
neighborhoods and distribution regions. As just one example of the
challenges, high penetration of uncoordinated rooftop solar can reduce
the reliability of distribution grids \cite{denholm2024maintaining}.

This paper introduces EcoNet, a Bayesian approach to household and
neighborhood energy management based on active inference \cite{SMITH2022102632}.
The EcoNet project aims to improve energy management and coordination,
while taking into account potentially conditional and conflicting
goals and preferences, and accommodating inherent uncertainties associated
with, for example, forecasts of weather, room occupancy, and energy
use and generation.

In general, conflicting goals can arise from many sources, including
the directive to conform to norms---legal, ethical, and cultural---which
themselves can conflict. In this sense, the EcoNet project addresses
artificial intelligence (AI) governance \cite{Dhaou-2024}, understood
as the mitigation of risks posed by AI systems. One such risk is the
failure of an artificial agent to comply with the complex knitting
of human-defined legal, ethical and cultural norms that underlies
human social coordination \cite{Constant-2025,Constant2025Path,Constant2024-ev}.

Here we present a simple, proof-of-concept model that involves energy
management in a single one-room household. The managed devices are
a HVAC system, a solar panel, and a battery that can be charged by
the grid or by solar. Devices are controlled by two agents. The discrete
actions available to the \textit{Thermostat Agent} are to set the
thermostat to \textit{cool}, \textit{heat}, or \textit{off}. The actions
available to the \textit{Battery Agent} are to set the battery to
\textit{charge}, \textit{discharge}, or \textit{off}. The household
prefers to minimize energy costs and grid-associated greenhouse gas
emissions, while preferences for room temperature are conditional
on ToU rates and expected room occupancy. By use of a multi-agent
approach, we set the stage for coordination in more complex settings,
such as between households and over larger regions.

The following sections discuss related work, provide an overview of
active inference, describe the EcoNet model, present simulation results,
and discuss outcomes and future work.

\section{Related Work}

Given the importance of distributed energy resources and stresses
on energy grids from increased electrification, a large body of research
addresses utility options related to DERs and energy use, including
utility-level control of or incentives for DER management \cite{smith-2024,Kurella31122025}.
Public datasets of household energy use and grid dynamics are available
to assist researchers \cite{Altamimi-2024,OEDI_Dataset_2981,NEEA_Org}.
From a utility perspective, households and commercial buildings can
offer flexibility services such as energy generation, load shifting,
and peak shaving, as well as ancillary services to help maintain grid
stability and function. These services include primary frequency response,
active power smoothing, exchange of reactive power for voltage regulation,
contributions to fault clearing, and voltage harmonic mitigation \cite{Demoulias-2020}.
Such capacities are of obvious interest to policymakers, planners,
manufacturers, consumer groups, and other stakeholders in energy and
environmental domains.

Utility programs that encourage households to offer flexibility and
ancillary services are typically implemented using price signals,
such as ToU rates, or other kinds of incentives. While price-based
programs empower prosumers and maintain privacy relative to direct
utility control of household devices, they depend on consumer attention
and action for success. Utilities thus benefit when households employ
Home Energy Management Systems (HEMS) for intelligent, automated or
semi-automated control of energy devices. Households benefit also,
as use of HEMS tends to reduce energy costs and increase comfort (e.g.,
via temperature control), among other benefits \cite{Mahapatra2022-zp}.

\subsection{RL and MPC for Home Energy Management Systems}

A sizable body of work addresses DER and energy management by HEMS
or related systems that serve building clusters or other localized
aggregations \cite{Zafar,Mahapatra2022-zp}. Households, neighborhoods,
and building clusters are the current focus of EcoNet.

Mason and Grijalva describe a reinforcement learning (RL) model applied
to building energy management \cite{mason2019review}. Wu et al. discuss
a deep-RL HEMS that uses intrusive load monitoring to recognize and
categorize human activity, the results of which are used to make energy-related
decisions \cite{WU2024114951}. Yu et al. describe a deep RL multi-agent
model that minimizes energy costs of HVAC systems in commercial buildings
\cite{yu2020multiagentdeepreinforcementlearning}. Touzani et al.
develop a deep RL approach for integrated control of HVAC and solar
battery storage and test it in a physical building against rule-based
strategies \cite{TOUZANI2021117733}. A variety of other groups have
studied RL frameworks for energy trading, demand response, load shedding,
voltage control, and microgrid dispatch \cite{shojaeighadikolaei2022distributed,shengren2023optimal,zhang2023deep,chen2022reinforcement,Ye-2023}.

Model predictive control (MPC), which typically uses rolling-horizon
optimization \cite{Schwenzer2021-bl}, has also been studied. Khabbazi
et al. review 24 field demonstrations of MPC and RL for HVAC control
in residential buildings and 80 in commercial buildings \cite{khabbazi2025}.
They report that average cost savings are 16 percent for residential
buildings and 13 percent for commercial buildings. Hartel and Bocklisch
compare RL and MPC systems for minimizing energy costs of solar battery
systems \cite{Hartel-2023}. Wang et al. compare several RL algorithms
against MPC and rule-based systems for heat pump control \cite{WANG2023120430}.

Despite their potential benefits neither MPC nor RL have seen wide
adoption in DER/energy management \cite{khabbazi2025}. Wang and Hong
identify three barriers for RL: 1) time-consuming and data-intensive
training requirements; 2) security and robustness concerns (regarding
the potential for detrimental or catastrophic decision making); and
3) lack of generalizability, where each new implementation requires
new training \cite{WANG2020115036}. A barrier for MPC is that it
requires a reasonably accurate model of the system and environment,
which is not always available, and which can change over time if devices
or patterns of management change.

Nweye et al. address the three RL barriers in their model of a 17-building
grid-interactive residential community \cite{NWEYE2023121323}. The
aim is to minimize energy consumption and cost, minimize emissions,
and provide grid flexibility services. To reduce data requirements
they implement digital twins of real buildings, constructed in the
CityLearn gymnasium \cite{Vazquez-2019}. They address security and
comfort via an occupancy-aware RL control architecture. And they address
generalizability through use of transfer learning.

In their system, the first step is to obtain a history of energy-related
data from real buildings. The second step is validation of the simulation
environment and control agent. Following validation, experiments are
carried out to determine what control policy has the best performance.
A best-performing control policy for a building is one that yields
the best outcome for a set of KPIs. Finally, the best-performing control
policies are deployed in the real-world grid-interactive community
in the respective building whose data they were trained on, or are
used as source for transfer learning and leveraged in other target
buildings.

In addition to the RL and MPC barriers already mentioned, DER systems
and environments can exhibit substantial uncertainty. Forecasts for
weather and solar generation can be inaccurate, for example. Conventional
RL and MPC models are not probabilistic, however, and so are not well
equipped to handle uncertainty. To address uncertainty quantification,
Zhang et al. develop a Bayesian deep RL control approach for microgrids
in which they replace the deterministic neural network in traditional
RL with a Bayesian probabilistic one \cite{zhang-202310005226}.

\subsection{Active Inference for Home Energy Management Systems}

In contrast to conventional RL and MPC, active inference employs a
generative probabilistic model that naturally accounts for uncertainty.
Like MPC, active inference requires a model, although that model can
be readily learned from data, especially if the model structure is
known. Unlike RL, active inference models are not usually data intensive,
and conditional constraints on behavior can be easily implemented.
Also unlike RL, a method for balancing agent exploration and exploitation
stems from first principles \cite{tschantz2020reinforcementlearningactiveinference}.
Furthermore, active inference is apt for multiple constraint satisfaction
problems, by specifying preferences in terms of probability distributions
over outcomes. Active inference is particularly useful when an agent
must plan ahead and can act to reduce uncertainty. For these reasons,
active inference offers an attractive approach for DER and energy
management. However, the study of active inference in this domain
is very new.

To the best of our knowledge, only a few studies exist on the topic.
We have explored the use of active inference in sustainability and
resource management more generally \cite{albarracin2024sustainability,albarracin2024modeling}.
Specific to HEMS, Nazemi et al. propose a dual-layer, hybrid active
inference architecture for tackling both building- and community-level
energy management \cite{danial2025active}. The bottom layer is a
continuous state-space model of building energy use, where HVAC systems
are controlled. The top layer (termed \textit{community manager})
is a discrete state-space model of aggregated energy use over buildings,
with the aim to keep power consumption within the proximity of day-ahead
plans. The community manager controls a shared energy storage system,
distributed renewables, and market transactions in response to external
price signals. Linking the two layers is a privacy-preserving interaction
protocol. In simulations, each building is represented by a digital
twin, an EnergyPlus model \cite{CRAWLEY2001319}, that integrates
real-time factors such as energy use, PV generation, energy storage,
and occupancy schedules.

The authors benchmark model performance against optimal planning (in
the fully observable case) and against RL approaches in which the
RL model is trained on a history of observables for the buildings
in question. Compared to RL models, the active inference agent predicts
indoor room temperatures that align more closely with desired temperatures.
This is not surprising, perhaps, as the active inference model encapsulates
prior knowledge about the HVAC and other systems. The community manager
successfully optimizes battery usage and market transactions, even
under an extreme pricing scenario.

Active inference applied to DER and energy management is new and promising,
potentially offering robust, uncertainty-aware coordination among
intelligent agents in a household and between households. Below, we
explore a multi-agent system in a single household. But before describing
that model, we provide background on active inference.

\section{Active Inference}

Active inference originated within the neuroscience field to explain
how agents perceive, learn, and act in uncertain environments. It
is a Bayesian model-based approach that considers perception, (parameter)
learning, and action as intertwined processes of inference \cite{friston2017graphical,millidge2020relationship,SMITH2022102632}.
Inference is achieved by minimizing free energy via variational inference.
As a brief primer, for data $x$ and a parameter $\theta$ treated
as a random variable, Bayes theorem is

\begin{equation}
p(\theta|x)=\frac{p(\theta,x)}{p(x)}=\frac{p(x\mid\theta)\,p(\theta)}{\int p(x\mid\theta)\,p(\theta)\D\theta}\,,
\end{equation}

where the marginal likelihood $p(x)$ is called the \textit{model
evidence}. The right-hand side is the \textit{likelihood} times \textit{prior,
$p(\theta)$,} divided by the evidence. The left-hand side is the
\textit{posterior,} which is the updated prior after seeing the data.
A typical approach to estimate $\theta$ is Maximum Likelihood Estimation
II:

\begin{equation}
p(\theta|x)=\underset{\theta}{\mathrm{arg\,max}}\int p(x\mid\theta)\,p(\theta)\D\theta\,.
\end{equation}

Commonly, the integral is intractable. Progress is made by maximizing
a lower bound called the \textit{variational lower bound} or \textit{evidence
lower bound} (ELBO). It can be derived by choosing a distribution
$q(\theta)$ that is similar to but simpler than $p(\theta\mid x)$,
and that by construction does not depend on $x$. Using Jensen's inequality
and after a series of manipulations, the result is

\begin{align}
\ln p(x) & \geq\mathbb{E}_{q}\left[\log\frac{p(\theta,x)}{q(\theta)}\right]\nonumber \\
 & =\mathbb{E}_{q}\left[\log p(\theta,x)\right]+H(\theta)\,,\label{eq:Jensen}
\end{align}

where the right-hand sides are called the ELBO, $H$ is entropy, and
the negative ELBO is called \textit{free energy:} 
\begin{align}
F(\theta) & =\mathbb{E}_{q}\left[\log q(\theta)\right]-\mathbb{E}_{q}\left[\log p(\theta,x)\right]\label{eq:FE}\\
 & \geq-\ln p(x)\,.\nonumber 
\end{align}
The right-hand side of the last line is called \textit{surprisal}.
Thus, minimizing free energy is equivalent to estimating the posterior
by minimizing an upper bound on surprisal.

By minimizing free energy, agents implicitly infer the hidden causes
of their sensory inputs and select actions that fulfill their goals
or objectives. In short, active inference unifies perception, learning,
and action selection under a single mathematical principle.

An active inference agent possesses (or is) a generative model of
its environment that predicts sensory inputs and the consequences
of actions. The agent updates its beliefs about the world in response
to observations and chooses actions that minimize expected free energy,
balancing exploration (reducing uncertainty) and exploitation (achieving
goals). The environment, in contrast, is a generative process that
reacts to agent actions, generating potentially noisy observations
that the agent must account for. Commonly, the agent is constructed
as a discrete, partially observed Markov decision process (POMDP),
although continuous state-space models are possible \cite{Wade-2021,Catal-2020,collis2024learninghybridactiveinference}.
Deep learning has also been explored in this setting as deep active
inference \cite{Ueltzhoffer2018,tschantz2019scalingactiveinference,champion2023deconstructingdeepactiveinference}.

In each time step of a simulation, an agent typically performs perception,
learning, policy evaluation, and action selection. Perception entails
inferring hidden states of the generative model based on potentially
noisy observations obtained from the environment. Learning involves
updating the parameters of the generative model to better explain
observed data. In policy evaluation, the expected free energy (EFE)
of each policy available to the agent is calculated. A policy is a
series of potential actions over a rolling horizon of a specified
length. Alternatively, marginal EFE over the next action can be calculated.
Action selection involves choosing a policy (or action) based on its
EFE. The selected action (e.g., the first action in a selected policy)
is then communicated to, and affects, the environment.

The core of active inference is a joint generative model,

\begin{equation}
p(\tilde{o},\tilde{s},\pi)\,,\label{eq:generative_model}
\end{equation}

where $o$ signifies observations; $s$ signifies states (beliefs
about hidden or latent states); the tilde accent signifies a series
of sequential values over a simulation, from time $t_{1}$ to $t_{T}$;
and $\pi$ signifies the policy that is chosen in each time step,
with each policy being a series of sequential actions over a policy
length, $\tau$.

The expected free energy is evaluated under a posterior predictive
density, $\bar{q}$, over observations in the future, conditioned
on a particular policy (with prior preferences or constraints usually
denoted by \textit{C}). The EFE is derived as follows: two variational
distributions are employed during simulation, both using the mean-field
approximation shown by terms to the right of the arrows:

\begin{align}
q(\tilde{s},\pi) & =q(\pi)q(\tilde{s}\mid\pi)\rightarrow q(\pi)\prod_{\tau}q(s_{\tau}\mid\pi)\\
\bar{q}(\tilde{o},\tilde{s},\pi) & =q(\pi)q(\tilde{o},\tilde{s}\mid\pi)\rightarrow q(\pi)\prod_{\tau}q(o_{\tau},s_{\tau}\mid\pi)\,.
\end{align}

The factorized distributions and the generative model can be substituted
into Equation \ref{eq:FE} to obtain the free energy for a particular
policy in a particular simulation step:

\begin{align}
F & =-\mathbb{E}_{q(\tilde{s}\mid\pi)}\left[\ln\frac{p(\tilde{o},\tilde{s}\mid\pi)\,p(\pi)}{q(\tilde{s}\mid\pi)\,q(\pi)}\right]\\
 & =-\mathbb{E}_{q(\tilde{s}\mid\pi)}\left[\ln\frac{p(\tilde{o},\tilde{s}\mid\pi)}{q(\tilde{s}\mid\pi)}\right]+D_{\mathrm{KL}}\left[q(\pi)\parallel p(\pi)\right]\nonumber \\
 & =\mathbb{E}_{q(\pi)}\left[F_{\pi}\right]+D_{\mathrm{KL}}\left[q(\pi)\parallel p(\pi)\right]\nonumber \\
F_{\pi} & =-\mathbb{E}_{q(\tilde{s}\mid\pi)}\left[\ln\frac{p(\tilde{o},\tilde{s}\mid\pi)}{q(\tilde{s}\mid\pi)}\right]\,.\nonumber 
\end{align}

To minimize $F$, we take its derivative with respect to $q(\pi)$
and set it to zero, resulting in

\begin{align}
q_{0}(\pi) & =\sigma\left(\ln p(\pi)-G_{\pi}\right)\\
G_{\pi} & =-\mathbb{E}_{q(\tilde{o},\tilde{s}\mid\pi)}\left[\ln\frac{p(\tilde{o},\tilde{s}\mid\pi)}{q(\tilde{s}\mid\pi)}\right]\,,
\end{align}

where $q_{0}$ is the belief prior to observations, $\sigma$ is the
softmax function, and the negative expected free energy, $G_{\pi}$,
is a sum over each step of a policy. The distribution $\bar{q}$ is
substituted for $q$ when taking expectations of free energy in the
equation for $G_{\pi}$ to finesse a shortcoming that would otherwise
occur. With $q$ rather than $\bar{q}$, $q_{0}$ would not accommodate
our prior knowledge that outcomes will become available in the future
\cite{parr2019}. With $\bar{q}$, future (not yet seen) observations
enter the expectations as random variables. In each simulation step
the agent selects the policy that it expects will lead to the lowest
free energy.

Dropping the accents, the negative expected free energy of a policy
step can be decomposed as

\begin{align}
G_{\pi} & =-\mathbb{E}_{q(o,s\mid\pi)}\left[\ln\frac{p(o,s\mid\pi)}{q(s\mid\pi)}\right]\nonumber \\
 & =\underset{\mathit{risk}}{\underbrace{D_{\mathrm{KL}}\left[q(o\mid\pi)\parallel p(o\mid C)\right]}}+\underset{\mathit{ambiguity}}{\underbrace{\mathbb{E}_{q(s\mid\pi)}\left[H(p(o\mid s)\right]}}\\
 & =-\underset{\mathit{information\,gain}}{\underbrace{\mathbb{E}_{q(o,s\mid\pi)}\left[\ln\frac{q(s\mid o,\pi)}{q(s\mid\pi)}\right]}}-\underset{\mathit{expected\,utility}}{\underbrace{\mathbb{E}_{q(o\mid\pi)}\left[\ln p(o\mid C)\right]}\,,}
\end{align}

where $p(o\mid\pi)$ has been replaced with $p(o\mid C)$, where $C$
is a vector of preferences. Thus, the agent selects actions that fulfill
goals (i.e., maximize expected utility) and that reduce uncertainty
(i.e., maximize information gain).

\section{Model Description}

Our HEMS model consists of two heterogeneous agents in a one-room
building. The energy-related devices are a HVAC system, a solar panel,
and a battery that can be charged by the grid or by solar. The building
also has a baseline pattern of energy use over the course of a day
(e.g. peak energy use in the morning and evening) that is not controllable.
The Thermostat Agent controls the thermostat, with discrete actions
of \textit{cool}, \textit{heat}, or \textit{off}. Its preference is
to keep the room at a comfortable temperature, conditional on occupancy
and ToU rate. If the room is expected to be unoccupied, a lower temperature
is preferred.

The Battery Agent controls the battery, with discrete actions of \textit{charge},
\textit{discharge}, and \textit{off}. Its preference is to minimize
total household energy costs and grid-associated greenhouse gas emissions.
Thus, not only do the two agents control different devices and consist
of different generative models, they potentially exhibit conflicting
preferences: a warm room, for the Thermostat Agent, and low energy
cost, for the Battery Agent.

\begin{figure}
\begin{centering}
\includegraphics[width=0.85\columnwidth]{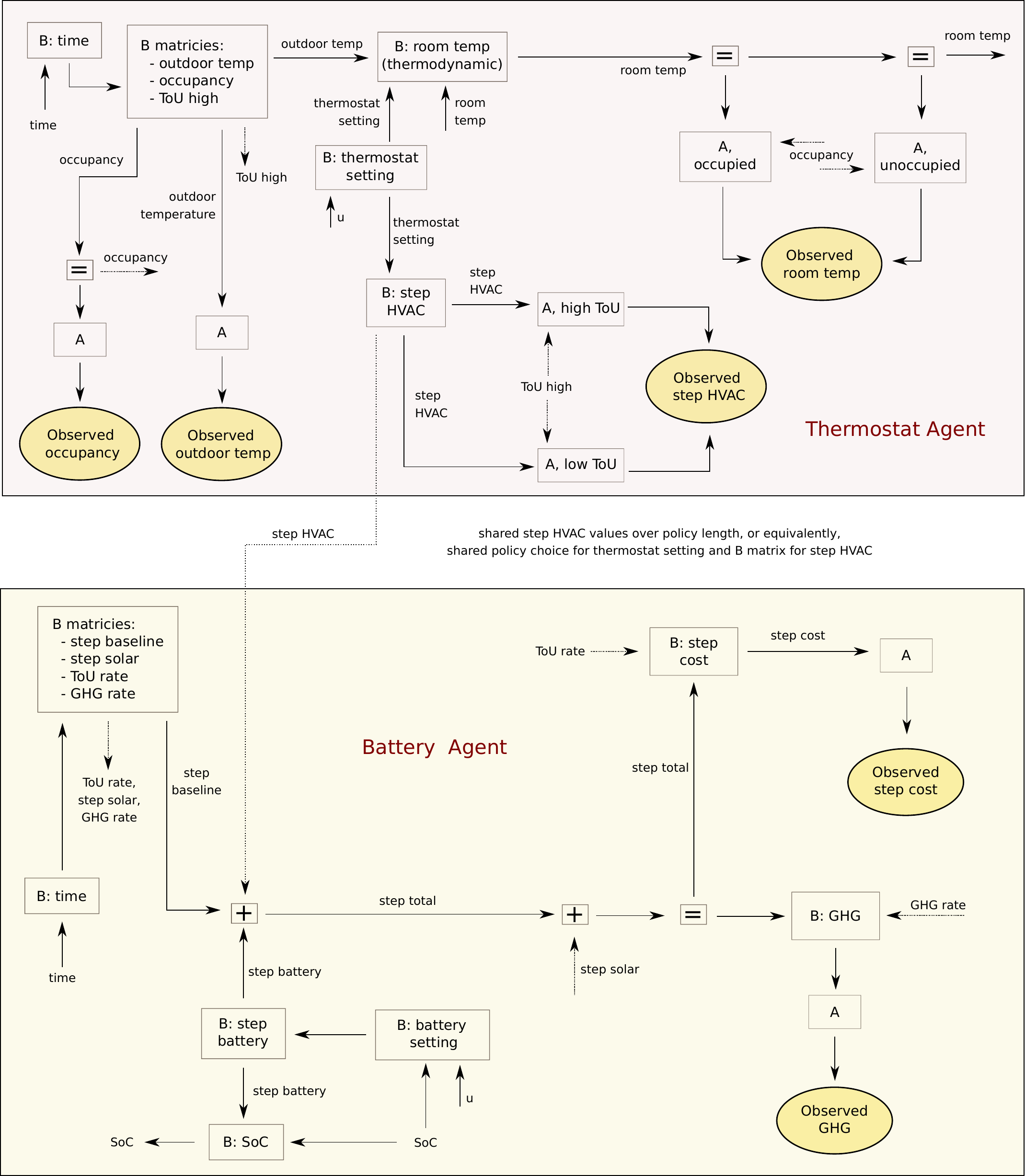} 
\par\end{centering}
\caption{Pseudo-factor graph for two-agent EcoNet model. See text for meaning
of state and observation variables and for information shared between
agents. Squares labeled $A$ and $B$ refer to $A$ (likelihood) and
$B$ (prior transition) matrices in the pymdp implementation. }\label{fig:factor=000020graph}
\end{figure}

The input data consists of a table of time series, based on two-hour
steps. Variables are time of day, solar generation, baseline load,
ToU rate, outdoor temperature, room occupancy, and greenhouse gas
emission rate. All values are synthetic, but are meant to reflect
semi-realistic conditions. Tracking and controlling the temperature,
energy use, solar generation, and other processes in a real building
is not the purpose of this study. Rather, the purpose is to demonstrate
proof of concept. A pseudo-factor graph depicting the model is provided
in Figure \ref{fig:factor=000020graph}.

In the figure, variables \textit{step\_{*}} (e.g., \textit{step\_battery},
\textit{step\_solar}) refer to kWh energy use or generation associated
with a particular state variable. Total energy use in a simulation
step (\textit{step\_total}) is the sum of baseline, HVAC, battery,
and solar energy, where solar generation and battery discharge are
negative energies. The variable \textit{step\_cost} is the total cost
of energy in a simulation step (\textit{step\_total} $\times$ ToU
rate). Each of the yellow ellipses signify an observable, and each
is associated with a preference (which can be flat). \textit{SoC}
is battery state of charge. \textit{GHG} is greenhouse gas emissions.
\textit{ToU\_high} is a Boolean variable that signifies whether or
not the ToU rate is high or low.

The $B$ (transition) matrix for room temperature encodes a simple
thermodynamic model, where new room temperature depends on previous
room temperature, outdoor temperature, thermostat setting, and the
size of the HVAC system. The $B$ matrices for HVAC energy, battery
energy, SoC, and total energy also encode simple formulas (for example,
the formula for total energy was just given). Updates for the other
state variables (e.g., ToU rate, outdoor temperature) are based on
look-ups in the input data table. Any degree of noise can optionally
be embedded in $A$ (likelihood) or $B$ matrices, in priors, and/or
in the underlying input data table. Parameter learning is also possible.

The discretisation levels for cost, temperature, and energy are set
by the user. Choices affect the size of the $A$ and $B$ matrices.
A very high discretisation level would approximate a continuous state-space
model, but would incur a memory and run-time penalty. The most important
factor influencing run time is the policy length. We found that a
policy length of six steps (12 hours) is needed to allow the agent
to adequately plan ahead for battery charge and discharge in our sample
problems. Other hyperparameters include target room temperature for
occupied and unoccupied status, policy length, initial beliefs about
room temperature and SoC, the simulation length, and whether parameter
learning is used.

In each simulation step, the Thermostat Agent evaluates and selects
a best policy. It passes the impacts of that policy on HVAC energy
use to the Battery Agent. The Battery Agent consider HVAC energy,
along with other energies, when it estimates total energy and total
cost.

The model is constructed in pymdp \cite{Heins_2022}. State transitions
are updated sequentially, such that the update of one state variable
can be used as input for updating other state variables. For example,
solar generation is updated based on lookup to the input data table.
The result of this update is used as input for updating total energy
(\textit{step\_total}). We used a modified version of pymdp that allows
this kind of cascading updates.

\begin{figure}
\begin{centering}
\includegraphics[width=0.85\columnwidth]{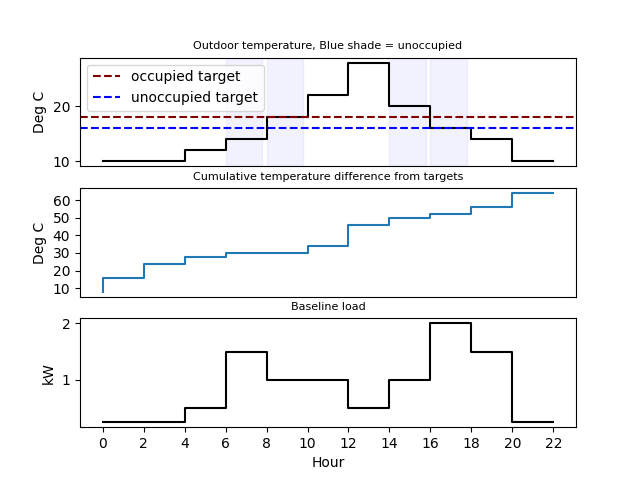} 
\par\end{centering}
\caption{Daily outdoor temperature and baseline energy use as examples of input
data (top and bottom panels). Shading indicates occupancy status.
Target temperatures for occupied and unoccupied status are prior preferences
set by the user. The middle panel shows cumulative absolute difference
between outdoor temperature and targets. This is the worst case scenario
for indoor room temperature, for a house without a HVAC system. }\label{fig:outdoor_temp}
\end{figure}

Figure \ref{fig:outdoor_temp} shows an example of input data, in
this case outdoor temperature and baseline energy use.

\section{Results}

The experimental setup allows for flexible changes to parameters,
hyperparameters, priors, and input data. This permits us to run simulations
under a variety of conditions, of different degrees of challenge.
In this section, we give representative results for two scenarios.
The first is a deterministic problem (zero uncertainty) where the
agents have accurate models of the environment. The policy length
is six steps (12 hours), and we also show results for a modified problem
that has four steps. Two days are simulated. The input data are the
same for each day. ToU rates are either high or low, and these are
associated with numerical values.

The results demonstrate that the agents try to achieve and maintain
target room temperatures, in spite of the under-powered HVAC system.
They also try to reduce costs and plan ahead to charge the battery
for later use.

The second scenario uses the same data and agents as the first, but
now the $B$ matrix for room temperature is fully uncertain (uninformative)
for the Thermostat Agent at the start of the simulation. As well,
the initial prior on room temperature is uninformative. The agent
must \textit{learn} the $B$ matrix that embeds the thermodynamic
model of the house, while \textit{inferring} the best course of action.

\subsection{Deterministic Scenario}

\begin{figure}
\begin{centering}
\includegraphics[width=0.7\columnwidth]{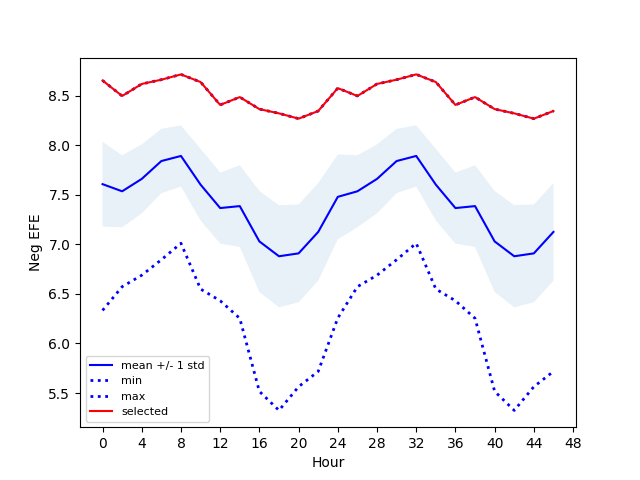} 
\par\end{centering}
\caption{Distribution of negative EFE over policies and simulation steps. }\label{fig:B-neg_efe}
\end{figure}

Figure \ref{fig:B-neg_efe} shows the distribution of (negative) expected
free energy (i.e., $G_{\pi}$) over all policies for the Battery Agent
(each agent evaluated 729 policies in each simulation step). Negative
EFE for the selected policy in each simulation step is shown in red,
which by design coincides with the maximum $G_{\pi}$ (policy selection
was deterministic). In each step, the agent takes the first action
in the selected policy. As shown, the EFE landscape varies across
simulation steps---the policies at each time step result in a range
of EFE values. Because EFE is the sum of information gain and expected
utility, a change in either across simulation steps can result in
policies that the agent has more or less commitment toward. By construction,
the average EFE corresponds to the entropy or uncertainty of (Bayesian)
beliefs about the policies that should be pursued. Fluctuations in
this uncertainty reflect the challenge of satisfying multiple constraints
or preferences in a context-sensitive fashion.

\begin{figure}
\begin{centering}
\includegraphics[width=0.85\columnwidth]{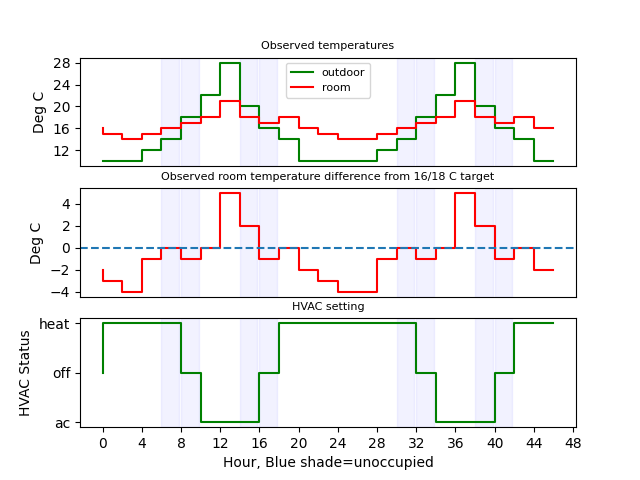} 
\par\end{centering}
\caption{Observed outdoor and room temperatures (top panel). Difference between
observed room temperature and targets (middle panel). The selected
thermostat actions (bottom panel). }\label{fig:A-temperature}
\end{figure}

Figure \ref{fig:A-temperature}, top panel, shows observed outdoor
and room temperatures, given the series of actions selected by the
agents. In a worst case scenario, for a house with no HVAC system,
the two curves would be superimposed. The middle panel shows the difference
between observed room temperature and target temperatures ($18^{\circ}$C
for occupied status, $16^{\circ}$C for unoccupied). The daily average
cumulative sum of absolute temperature differences from the targets
is 22 degrees C, which is less than the worst case of 64 degrees,
shown in Figure \ref{fig:outdoor_temp}. The bottom panel shows the
selected thermostat actions.

The Thermostat Agent tends to turn on the heat when the room is below
the target temperatures and turn on the air conditioning (ac) when
the room is above the target temperatures. Its actions cannot maintain
target room temperatures, however, because the HVAC system is underpowered.
We designed the HVAC system this way to challenge the agent. As would
be expected, the largest differences between targets and room temperature
occur during cool nights and warm afternoons.

\begin{figure}
\begin{centering}
\includegraphics[width=0.7\columnwidth]{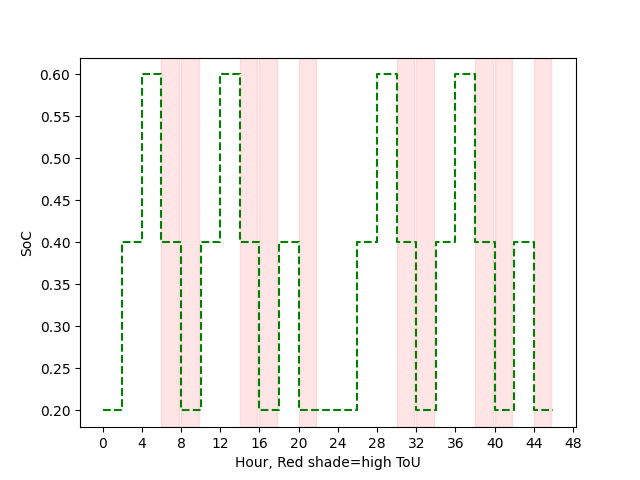} 
\par\end{centering}
\caption{Agent beliefs about battery state of charge. Beliefs about ToU rates
are shaded. }\label{fig:B-soc-vs-tou}
\end{figure}

The Battery agent has a preference to minimize energy costs. One way
it can do so is to discharge the battery during high-ToU periods,
when energy is expensive. Discharged energy is used by the household,
if needed, and any excess is sold back to the grid. Figure \ref{fig:B-soc-vs-tou}
shows that the agent tends to use the battery (indicated by a drop
in SoC) during high-ToU periods (red shading). Similarly, it tries
to charge the battery during low-ToU periods.

Battery SoC is discretized into five categories, from 0.0 to 1.0 in
steps of 0.2. To protect the battery, the Battery Agent is not allowed
to discharge when SoC $\leq$ 0.2 or to charge when SoC $\geq$ 0.8.
Each two-hour step can charge or discharge the battery 20 percent.
The initial charge is 0.2. Starting from an SoC of 0.2, the agent
can maximally charge the battery in three consecutive two-hour blocks.
It never needs to charge the battery to a SoC of 0.8, however, because
all high ToU periods last for no more than two steps. The charges
it attains are sufficient to cover all high-ToU periods.

\begin{figure}
\begin{centering}
\begin{tabular}{|c|c|}
\hline 
\includegraphics[width=0.45\columnwidth]{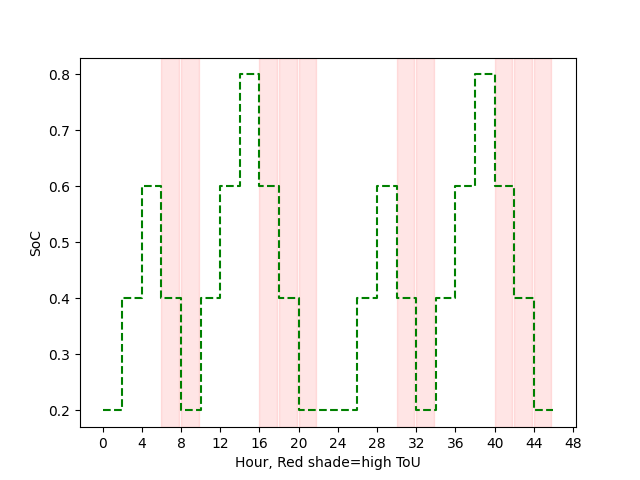}  & \includegraphics[width=0.45\columnwidth]{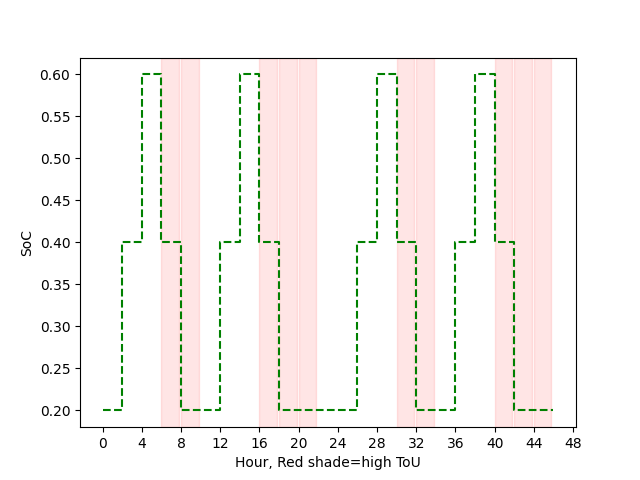}\tabularnewline
\hline 
\end{tabular}
\par\end{centering}
\caption{Agent beliefs about battery state of charge when an evening high-ToU
period last for three hours. Policy length is six steps (left panel)
and four steps (right panel). Beliefs about ToU rates are shaded.
}\label{fig:variation}
\end{figure}

As a variation of this problem, we alter the ToU schedule so that
during each evening there is a three-hour high-ToU period. The policy
length is again six steps. Now the agent charges the battery fully
(to a SoC of 0.8) so that it can discharge during all high-ToU steps.
Results are shown in Figure \ref{fig:variation} (left panel). As
another variation, we run the same simulation but this time using
a policy length of four steps (eight hours). Now the agent cannot
see far enough into the future to perfectly plan ahead for the three-hour
high-ToU period. It starts charging too late, and as a result cannot
discharge during all high-ToU steps.

\begin{figure}
\begin{centering}
\includegraphics[width=0.7\columnwidth]{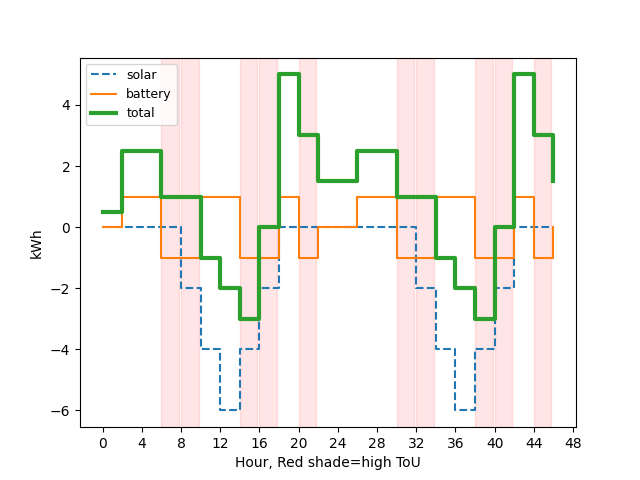} 
\par\end{centering}
\caption{Agent beliefs about total, battery, and solar energy use. }\label{fig:B-step-cost-vs-tou}
\end{figure}

Figure \ref{fig:B-step-cost-vs-tou} shows total energy use in the
house during each step of the original problem, along with solar generation
and battery charge and discharge. HVAC and baseline energy are not
shown. Battery charge registers as a positive energy, and discharge
and solar generation as negative energy. The four largest positive
peaks in total energy roughly coincide with morning and evening peak
baseline energy use, but are offset by the agent to occur during low-ToU
periods. The two largest negative peaks roughly coincide with solar
generation, but are offset to occur during high-ToU periods. The battery
only charges during low-ToU periods and only discharges during high-ToU
periods. Charging also tends to coincide with solar generation.

\begin{figure}
\begin{centering}
\includegraphics[width=0.85\columnwidth]{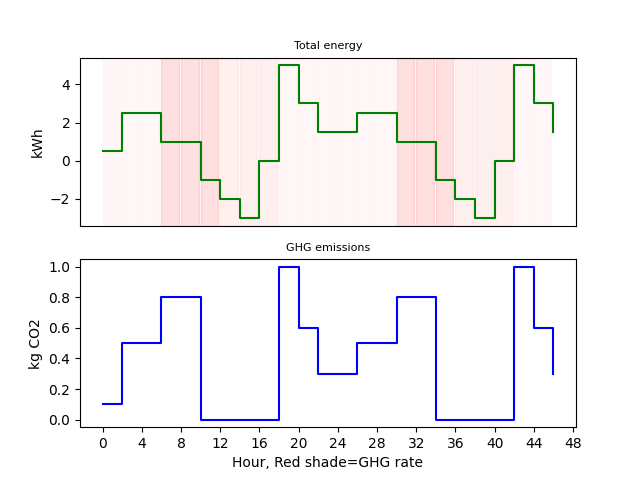} 
\par\end{centering}
\caption{Agent beliefs about total energy use versus grid-associated GHG rates
(top panel) and emissions due to energy use (bottom panel). Deeper
red shading indicates higher GHG rates. }\label{fig:B-ghg}
\end{figure}

Figure \ref{fig:B-ghg} shows agent beliefs about total energy use
in comparison to grid-associated GHG rates (top panel). Peak energy
use is offset by the agent to avoid high-GHG rate periods. The bottom
panel shows beliefs about GHG emissions due to energy use.

\subsection{Parameter Learning Scenario}

\begin{figure}
\begin{centering}
\begin{tabular}{|c|c|}
\hline 
\includegraphics[width=0.45\columnwidth]{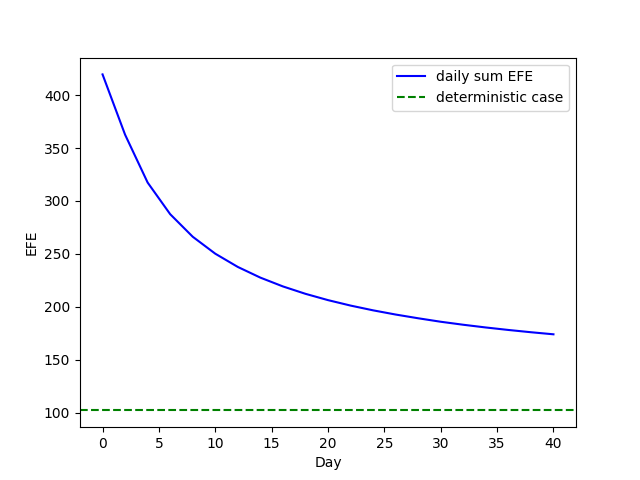}  & \includegraphics[width=0.45\columnwidth]{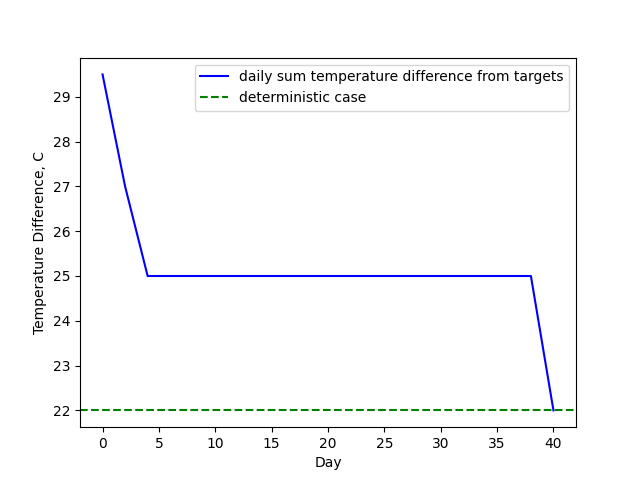}\tabularnewline
\hline 
\end{tabular}
\par\end{centering}
\caption{Learning the room temperature transition matrix. Drop in negative
expected free energy over a 40-day training period (left panel). At
40 days the agent's ability to achieve target room temperature is
the same as seen in the deterministic model (right panel). }\label{fig:learning}
\end{figure}

The second scenario investigated is identical to the deterministic
case, but now the Thermostat Agent starts with an uninformative transition
matrix for room temperature, and an uninformative belief about initial
room temperature. In effect, the agent needs to learn the thermodynamic
model of the house. The 24-hour input data are repeated in every day
of the trial. Figure \ref{fig:learning} shows that negative expected
free energy drops during learning, as the agent becomes more confident
about inferred policies (left panel), and that by day 40 the Thermostat
Agent is able to achieve deviations from target room temperature equal
to those of the deterministic model (right panel). The blocky nature
of the right panel is due to the discretisation level and low variability
in the (daily repeated) input data. In a realistic setting, some information
about the thermodynamic model would likely be available, and if so
the transition matrix might need to be refined, rather than learned
from scratch.

\section{Discussion}

As societal electrification increases in response to climate change,
and as the penetration of DER devices in distribution grids increases,
it becomes increasingly important for grid flexibility and stability
not only that each energy device is intelligently managed, but that
management is coordinated between devices within a household, between
households of a neighborhood or complex, and across regions of the
grid. Intelligent management involves forecasts (of weather, solar
generation, energy use, household occupancy, etc.) and such forecasts
involve uncertainty. Related, the thermodynamic model of a house is
typically unknown and must be estimated, which also involves uncertainty.
Thus, the use of multi-agent probabilistic models for HEMS appears
promising. In this, the use of active inference seems particularly
promising, as an active inference agent can choose actions to resolve
uncertainty. In an advanced model for example, an agent might ask
a householder if a room will be occupied later in the evening, if
it is unsure. Relatively few HEMS studies conducted so far consider
probabilistic models, however.

Here we demonstrate a proof-of-concept active inference model. Results
suggest that the model is able to plan ahead to balance conflicting
preferences for energy use, energy cost, greenhouse gas emissions,
and comfort, as well as to learn parameters of an under-specified
house thermodynamic model.

Multiple studies suggest that HEMS and related energy management systems
have potential to reduce energy costs, conserve resources, reduce
greenhouse gas emissions, and increase grid stability and function
\cite{khabbazi2025,Mahapatra2022-zp,Martinez-2021,Tuomela-2021}.
To highlight one example, Gualandri and Kuzior \cite{Gualandri-2023}
demonstrate that increased adoption of HEMS could dramatically reduce
residential-energy consumption in Spain, which in their view could
help the country to achieve 2030 European energy targets. The degree
of future HEMS adoption depends of course on choices made by manufacturers,
utilities, consumer groups, end users, policymakers, and other stakeholders,
which in turn depends on the degree of cooperation between these entities
\cite{oluokun-2025,land14030566,Khafiso-2024}.

There are many possibilities for future work that could extend or
improve upon the current model. Like Nazemi et al. we could employ
a continuous state space and test the model using data generated from
real buildings or accurate digital twins. Although our aim is to coordinate
energy management across households, the current model only considers
management within a single building. Extension to neighborhoods or
building clusters would naturally expand our two-agent model to a
many-agent approach. Here, we use deterministic forecasts of weather,
solar generation, and other factors, but future efforts could use
probabilistic forecasts. More efficient methods for parameter and
structure learning are also possible, especially for learning complex
thermodynamic models of buildings. Renormalization group models or
tensor networks could be interesting here due to their capacity to
capture scale-invariance and compress data associated with high dimensional
functions, including probability functions \cite{Friston-2025,Roa-2024}.
Finally, it would be interesting to apply an active inference model
to infer the goals, preferences, and beliefs of households that largely
determine how they manage their energy devices. For example, a solar
battery inverter can be managed in numerous ways, for numerous purposes,
each of which could alter the voltages, currents, and power factors
seen at the point of common coupling with the grid. Given a history
of device management, as evidenced by the numerical readings available
to grid operators, what model best explains the management practices
of a household, and so also its expected energy impacts on the grid?

\section{Declarations}

\subsection{Availability of Data and Material}

The data used in this paper are synthetic examples and will not be
shared.

\subsection{Competing Interests}

The authors declare the following financial interests/personal relationships
which may be considered as potential competing interests: All authors
report a relationship with VERSES that includes employment or contract
agreement.

\subsection{Funding}

Funding for this study was provided by VERSES.

\subsection{Authors' Contributions}

All authors participated in manuscript preparation. All authors except
Esterhuysen participated in conceptual design of the model. Boik,
Esterhuysen, Hynes, and Kiefer participated in coding and simulation
of the model.

\subsection{Acknowledgments}

The authors would like to thank VERSES for its financial support of
this project.

 \bibliographystyle{plain}
\bibliography{references}

\end{document}